\vglue 12mm

\vskip 100pt
\centerline{\bf \
{ON THE SUPER FIELD REALIZATION OF SUPER CASIMIR
         ${\cal WA}_{n}$ ALGEBRAS
 }}
\vskip 5mm
\centerline{H.\ T.\ \"OZER \footnote{$^{^{_{\dagger }}}$}
{e-mail\ :\ ozert @ itu.edu.tr }}

\vskip 5mm
\centerline{\it {Physics Department,\ Faculty of Science and Letters,
Istanbul Technical University
,}}
\centerline{\it { 80626,\ Maslak,\ Istanbul,\ Turkey }}
\vskip 15mm
\noindent

We give an explicit quantum super field construction of the
$\cal{N}$=2 super- Casimir ${\cal WA}_{n}$ algebras, which is
obtained from supersymmetric Miura transformation for the
Lie superalgebra ${\cal A}_{n,n-1}$.And also we give an
extension of this algebras including a super vertex operator
which depends on simple root system of ${\cal A}_{n,n-1}$.

\vskip 30mm

\par\vfill\eject


\noindent{\bf{1.\ Introduction}}
\vskip5mm
\noindent Superconformal symmetry  has played an important role in the
developments  of the many branches of theoretical physics and mathematics.
Extended superconformal symmetry is an extension of superconformal symmetry
containing higher conformal spin generators in addition to the usual
superconformal algebra generators . Many superconformal algebras are known
 (for review see Ref.1 and referances therein).It is well known that a usual
 and a  fermionic ($\cal{N}$=1) Casimir algebra corresponding
to a simple Lie algebra $\cal{L}$${_n}$ can be denoted by $\cal{WL}$${_n}^{2-10}$.
Analogous, a ($\cal{N}$=2)  super Casimir algebra associated  with  the Lie superalgebra
${\cal A}$ $_{n,n-1}$ can be also shown by super-$\cal{WL}$${_n}^{11-13}$. Therefore
super-Casimir ${\cal {WA}}$$_{_n}$ algebras based on the Lie superalgebras ${\cal A}_{n,n-1}$.
With the above identification, one says that ${\cal {WA}}$$_{_2}$  algebra is the simplest
($\cal{N}$=2 ) superconformal algebra.\ This superconformal algebra coincides with Neveu-
Schwarz algebra ($\cal{N}$=1 super Virasoro algebra).\ We must emphasize here that the
($\cal{N}$=1) super- Casimir algebra is a subalgebra of the ($\cal{N}$=2) super-Casimir algebra.
L.J.Romans have proven that super- ${\cal {W}}$$_{_3}$ algebras $^{11}$ are associative for all
values of central charge c. Some realizations of this algebra in terms of the ($\cal{N}$=2)
superfields has been constructed with  (${n}$=2) two scalar fields by using the
supersymmetric Miura transformation $^{10,15,16}$

In this letter, we study a super Casimir algebra for a  Lie superalgebras
$\cal{L_{_N}}=\cal{A}$$_{n,n-1}$ ,\ which are called super-Casimir ${\cal {A}}$$_{_n}$
algebras,\ for  $n=3$ . We must emphasize here that all pure conformal fields
depand on n, so this is the first pure spin field definitions for this algebra in the
literature. We reserve a notation of our spin content as module-1\ : \
 $\{1,+{3\over 2},-{3\over 2},2\}$ and module-2\ : \ $\{2,+{5\over 2},-{5\over 2},3\}$.

Let us now give outline of this paper. In Sec. 2,\ we give a basis for
super-Casimir ${\cal {A}}$$_{_n}$ algebra by using the well-known supersymmetric Miura
transformation with $2 n$ free massless bosonic fields and  real fermion fields$^{12-16}$.\
Then,\ we constructed a free field realization of the super-Casimir ${\cal {A}}$$_{_n}$
algebra by calculating all primay fields explicitly in general n ,but we can not give all
nontrivial OPEs among primary fields.\ since all OPEs exist in the literature with some
normalizations.\ In Sec.3,\ we constructed, in the component language, a super- vertex
operator  extension of the super-Casimir ${\cal {A}}$$_{_n}$ algebras by
calculating explicitly nontrivial OPEs between  primary
fields and super-vertex operators which are of two forms.\ The calculations of
OPEs have been done with the help of Mathematica Package OPEDefs.m of
Thielemans $^{17}$ under the {\bf{Mathematica$^{TM}$}} in Ref.18.

\vskip5mm

\noindent{\bf{ 2. \ Super- Miura Basis For Super Casimir
${\cal WA}_{n}$ Algebras In the Component Formalism}}
\vskip 5mm

\noindent In this section, we begin with the review of the supersymmetric
Miura transformation based on a Lie superalgebra $\cal{A}$$_{n,n-1}$$^{12-16}$.
We use odd simple root system $ \{ \alpha_1,\cdots, \alpha_{2n} \} $ satisfying Cartan matrix
$C_{ij}\ =  \alpha_i \ .\ \alpha_j  =\big(-1 \big)^{i-1} \delta_{i+1,j}$ and
 $ \{ \lambda_1,\cdots, \lambda_{2n} \} $ are the fundamental weights of
${\cal A}$ $_{n,n-1}$ , satisfying $ \alpha_i \ .\ \lambda_j  =\delta_{ij}$. We
show ($\cal{N}$=1) super coordinate ${\cal Z}=\big(z,\theta \big)$ ana a super derivative
${\cal D}={\partial \over{\partial\theta}}+\theta {\partial \over{\partial z}}$.Then
 we introduce a primary basis for the the super-Casimir ${\cal {WA}}$$_{_n}$
algebras from $2 n$ free massless bosonic fields $\varphi_i(z)\ (i=1 , \cdots , 2 n)$
and  real fermion fields $\psi_i (z)\ (i=1 , \cdots , 2 n)$ realization point of view,
which are single-valued functions on the complex plane and its mode expansion are given by

$$
i\,\partial{\varphi}(z)=\sum_{n \in Z} a_n z^{-n-1},\ \ \ \ \ \
\psi(z)=\sum_{n \in Z} \psi_n z^{-n-{{1}\over{2}}}.
\eqno(2.1)
$$
respectively.\ A scalar superfield can be defined by
 $ \Phi \big({\cal Z} \big)\ = \ \varphi (z)\ + i\ \theta \ \psi (z)$. Canonical quantization
 gives the commutator and anti-commutator relations
$$
[a_m,a_n]=m \delta_{_{m+n,0}}\ , \ \ \ \ \ \ \
\{\psi_m,\psi_n\}=\delta_{_{mn}}\ ,
\eqno(2.2)
$$
and these  relations are equivalent to the contractions,\ as $z_{12}=z_{1}-z_{2}$
$$
\partial\varphi(\underline{z_1)\,\partial\varphi(}z_2)=\ {{1}\over {z_{12}^2}}
,\ \ \ \ \ \ \
\psi(\underline{z_1)\,\psi(}z_2)={{1}\over {z_{12}}}
\eqno(2.3)
$$
respectively.

\par\vfill\eject

Let us consider the supersymmetric Miura transformation $^{12-16}$, which is defined as

$$
{\cal R}_n ({\cal Z})={\cal D}^{2 n+1}\ +\  \sum_{j=2}^{2 n+1}\ {\cal W}_{i\over 2}({\cal Z})
\big( \alpha_{_0} \ {\cal D} \big)^{2 n+1-i}
$$
$$
=: \prod_{j=1}^{2 n+1} \big(\alpha_{_0} {\cal D}- \Theta_{_j}({\cal Z})\big):,
\eqno(2.4)
$$
where $ \Theta_{_j}\big( {\cal Z} \big)\  = \   \big(-1 \big)^{i-1} \  \big(\lambda_i-\lambda_{i-1} \big)\ .  \
{\cal D}\Phi \big({\cal Z} \big)\  \big({\lambda_0=\lambda_{2 n+1}=0} \big)   $
the symbol $ : ~* ~:$  shows the normal ordering,\ and $\alpha_{_0}$ is a free parameter.

The super- Casimir ${\cal {WA}}$$_{_n}\  $  algebra generated by a set of
chiral currents $  {\cal W}_{i\over 2}({\cal Z})$ \ ,\ of conformal dimension ${i\over 2}$~$~
(i=2 , \cdots , 2 n+1) $\. \ We show first few examples of currents, under
$ {\cal H}_{_i} = \lambda_i\ .\ {\cal D}\ \Phi \big({\cal Z} \big)$\  definition\ :

$$
{\cal W}_{_1}({\cal Z})=
-\,\sum_{i=1}^{2 n-1}\,({\cal H}_{_i}\,{\cal H}_{_{i+1}})({\cal Z})
-a_0\,\sum_{i=1}^{2 n}(-1)^i \,({\cal H}_{_i})({\cal Z})
\eqno(2.5)
$$
$$
{\cal W}_{_{3\over2}}({\cal Z})=
-\,\sum_{i=1}^{n}\,({\cal D} {\cal H}_{_{2i-1}}\,{\cal H}_{_{2 i}})({\cal Z})
+\,\sum_{i=1}^{n-1}\,({\cal H}_{_{2i}}\,{\cal D} {\cal H}_{_{2 i+1}})({\cal Z})
-a_0\,\sum_{i=1}^{n}\,({\cal D}^2 {\cal H}_{_i})({\cal Z})
\eqno(2.6)
$$
\noindent In the component formalism  all bosonic and fermionic currents 
 are expressed as in module-(k-1)\ :

$$
{\cal W}_{_{k-1}}({\cal Z})={\cal J}_{_{k-1}}({z})+\ i  \ \theta\
\bigg{[} {\cal G}_{_{k-{1\over2}}}^{+}({z})+{\cal G}_{_{k-{1\over2}}}^{-}({z}) \bigg{]}\
$$
$$
{\cal W}_{_{k-{1\over2}}}({\cal Z})= \alpha_{_0} \
\bigg{[}\ i \ {\cal G}_{_{k-{1\over2}}}^{-}({z})+\ \theta\ {\cal T}_{_{k}}({z}) \bigg{]}\
\eqno(2.7)
$$
For the module-2
$$
{\cal J}_{_{1}}({z})=
\,\sum_{i=1}^{2 n-1}\,(\psi_{_i}\,\psi_{_{i+1}})(z)
-a_0\,\sum_{i=1}^{2 n}(-1)^i \,\psi_{_i}(z)
$$
$$
{\cal G}_{_{{3\over2}}}^{-}({z})=
\,\sum_{i=1}^{2 n-2}\,(1-t_{_i})(\psi_{_i}\,h_{_{i+1}})(z)
-\,\sum_{i=1}^{2 n}\,t_{_i}(h_{_i}\,\psi_{_{i+1}})(z)
-a_0\,\sum_{i=1}^{2 n}(-1)^i \,(1-t_{_i})\psi_{_i}'(z)
$$
$$
{\cal G}_{_{{3\over2}}}^{+}({z})=
-\,\sum_{i=1}^{2 n-2}\,(1-t_{_i})(h_{_i}\,\psi_{_{i+1}})(z)
+\,\sum_{i=1}^{2 n}\,t_{_i}(\psi_{_i}\,h_{_{i+1}})(z)
+a_0\,\sum_{i=1}^{2 n}(-1)^i \,t_{_i}\psi_{_i}'(z)
$$
and
$$
{\cal T}_{_{2}}({z})=
\,\sum_{i=1}^{2 n-1}\,(-1)^i (h_{_i}\,h_{_{i+1}})(z)
+\,\sum_{i=1}^{2 n-1}\,(1-t_{_i})(h_{_i}\,h'_{_{i+1}})(z)
+\,\sum_{i=1}^{2 n-1}\,t_{_i}(h'_{_i}\,h_{_{i+1}})(z)
-a_0\,\sum_{i=1}^{2 n}(-1)^i (1-t_{_i})\,h_{_i}'(z)
\eqno(2.8)
$$
where $ {h}_{_i}(z) =\lambda_i . {\partial} \varphi(z) $\ ,
  $t_{_i}=\cases {1 \ \ & odd \ i\cr0\ \ & even \ i \cr}$
and also module-3 components including higher order
${\cal J}_{_{2}}({z})$,\
${\cal G}_{_{{5\over2}}}^{-}({z})$,\
${\cal G}_{_{{5\over2}}}^{+}({z})$ and
${\cal T}_{_{3}}({z})$
currents are not given  here since their formal complexity.$ {h}_{_i}(z)$'s and ${\psi}_{_i}(z)$'s
satisfy
$
h_{_i}(\underline{z_1)\,h_{_j}(}z_2) = {{[C^{-1}]_{ij}}\over {{z_{12}}^2}}
$
and
$
\psi_{_i}(\underline{z_1)\,\psi_{_j}(}z_2)={{[C^{-1}]_{ij}}\over {z_{12}}}
$
OPE's respectively, under (2.3) contractions ,
where $[C^{-1}]_{ij}$ is inverse matrix of Cartan matrix $C_{ij}$ of
Lie superalgebra $\cal{A}$$_{n,n-1}$.\

\par A definition
$
\top (z)\ =\ {\cal T}_{_2}(z)-{1\over2}\ {\cal J}_{_1}'(z)
$
denotes the energy-momentum tensor of $\cal{N}$=2 model with the central charge
$c^{_{({\cal{N}}=2)}}=3\ n \ \big( 1 - (n + 1 )a_{_0}^2\big)$ . The OPE with itself is

$$
\top(z_1)\top(z_2)=
{c^{_{({\cal{N}}=2)}}/2 \over z_{12}^4}
+{{2 \top(z_2)} \over z_{12}^2}+
{{\partial{\top(z_2)}} \over z_{12}}+\cdots
\eqno(2.9)
$$
Note that in the component definitions (2.7),all bosonic and fermionic currents are not primary
fields with respect to the energy-momentum tensor $\top (z)$, except ${\cal{J}}_{_1}(z)$, for bosonic
currents,\ i.e.

$$
\top (z_1){\cal{J}}_{_k}(z_2)={1 \over 2} \sum_{s=2}^k {(n - k + s)! \over (n - k)!}
2^{ s - 2}\ a_{_0}^{2 s  -  2}\Big(  1  +  (n  +  1  +  2(k  -  s))a_{_0}^2
\Big) {{{\cal{J}}_{_{k  -  s}}(z2)} \over z_{12}^{s  +  2}}
$$

$$
+{(n  - k  +  1)!  \over (n  -  k)! }
2^{ k  -  2}  a_{_0}^2 {{{\cal{J}}_{_{k  -  1}}(z_2)} \over z_{12}^3}
+{{k {\cal{J}}_{_k}(z_2)} \over z_{12}^2}+
{{\partial{{\cal{J}}_{_k}(z_2)}} \over z_{12}}+\cdots
\eqno(2.10)
$$
and  also fermionic currents, i.e.

$$
\top (z_1){\cal{G}}_{7\over 2}^{+}(z_2)=
(2 n-5) a_{_0}^2 \Big(  1  +  (n  + 3)a_{_0}^2\  \Big) {{{\cal{G}}_{_{3\over 2}}^{+}(z_2)} \over z_{12}^{4}}+
(2 n-2) a_{_0}^2\ {{{\cal{G}}_{_{5\over 2}}^{+}(z_2)} \over z_{12}^{3}}+
$$
$$
{{{{7\over 2}\ \cal{G}}_{_{7\over 2}}^{+}(z_2)} \over z_{12}^{2}}+
 {{{\partial \cal{G}}_{_{7\over 2}}^{+}(z_2)} \over z_{12}}+\cdots
\eqno(2.11)
$$
In order to construct all pure bosonic and fermionic currents for  the  $\cal{N}$=2 super- Casimir
${\cal WA}_{n}$ algebras, one may calculate suitable coefficients for quasi-primary bosonic and
fermionic fields ,which form a ${\cal N}=2$ super multiplet
${\kappa}
\{
{\cal J}_{_{2}}({z}),\
$
$
{{1}\over{\sqrt 2}}{\cal G}_{_{{5\over2}}}^{\mp}({z}),\
{\cal T}_{_{3}}({z})
\}
$
$$
\kappa\
\tilde{{\cal{J}}}_{_2}(z)=
 {\cal{J}}_{_2}(w)
-{1\over 2} (n - 1) a_{_0}^2\ \partial {\cal J}_{_1}(z)
-{(n - 1)\big(3+(2+3 n)a_{_0}^2\big)\over
     6 n-2+6 n (n+1)a_{_0}^2}\ ({\cal{J}}_{_1}  {\cal{J}}_{_1})(z)
$$
$$
+{(n - 1)\big(1-n a_{_0}^2\big) \big(1+(n+1) a_{_0}^2\big)\over
 3 n-1+3 n (n+1)a_{_0}^2}\top(z)
$$
$$
{\kappa\over{\sqrt 2}}\
\tilde{{\cal{G}}}_{_{5\over 2}}^+(z)={{\cal{G}}}_{_{5\over 2}}^+(z)
+{(n - 1)\big(1-n a_{_0}^2\big) \big(1+(n+1) a_{_0}^2\big)\over
 3 n-1+3 n (n+1)a_{_0}^2}\partial {{\cal{G}}}_{_{3\over 2}}^+(z)
-{(n - 1)\big(3+(2+3 n)a_{_0}^2\big)\over
     3 n-1+3 n (n+1)a_{_0}^2}\ ({\cal{J}}_{_1} {{\cal{G}}}_{_{3\over 2}}^+ )(z)
$$
$$
{\kappa\over{\sqrt 2}}\
\tilde{{\cal{G}}}_{_{5\over 2}}^-(z)={{\cal{G}}}_{_{5\over 2}}^-(z)
-{(n - 1)\big(1+3 n a_{_0}^2+2 n (n+1) a_{_0}^4\big)\over
 3 n-1+3 n (n+1)a_{_0}^2}\partial {{\cal{G}}}_{_{3\over 2}}^-(z)
-{(n - 1)\big(3+(2+3 n)a_{_0}^2\big)\over
     3 n-1+3 n (n+1)a_{_0}^2}\ ({\cal{J}}_{_1} {{\cal{G}}}_{_{3\over 2}}^-)(z)
$$
$$
\kappa\
\tilde{{\cal{W}}}_{_3}(z)={{\cal{T}}}_{_3}(z)-{1\over 2} \partial {\cal J}_{_2}(z)
-{(n - 1)\big(3+(2+3 n)a_{_0}^2\big)\over
     3 n-1+3 n (n+1)a_{_0}^2}\ \Big[ ({\cal{G}}^+_{_{3\over 2}}
{{\cal{G}}}_{_{3\over 2}}^-)(z)-({\cal{J}}_{_1} {{\cal W}_2})(z)
+{1\over 4}  \partial ({\cal{J}}_{_1}  {\cal{J}}_{_1})(z)\Big]
$$
$$
-{(n - 1)\big(3+(6 n+1) a_{_0}^2+3 n (n+1) a_{_0}^4\big)\over
 6 n-2+6 n (n+1)a_{_0}^2}\partial {\top}(z)-
{(n - 1)\big(1+n a_{_0}^2\big) \big(2+(n+1) a_{_0}^2\big)\over
 12 n-4+12 n (n+1)a_{_0}^2}\partial^2{\cal J}_1(z)
\eqno(2.12)
$$
where
$$\kappa\ = \ \sqrt
{{3 n-1 +3 n (n+1) a_{_0}^2}\over
{\big(1 - n a_{_0}^2\big) \big(2 + (n+1) a_{_0}^2\big) \big(1 + (n+2) a_{_0}^2\big)}}
\eqno(2.13)
$$
From (2.12), we may compute   the operator product expansions for the $\cal{N}$=2 super multiplet
${\kappa}
\{
{\cal J}_{_{2}}({z}),\
$
$
{{1}\over{\sqrt 2}}{\cal G}_{_{{5\over2}}}^{\mp}({z}),\
{\cal T}_{_{3}}({z})
\}
$.\
It is easy to see that these OPE's coincide with the results of the OPE method $^{11}$, under (2.12) normalization.

\vskip5mm

\noindent{\bf{ 3.\ (OPEs) for Chiral Super-Vertex Operators}}
\vskip 5mm
\noindent
In this section we define a super- vertex operator$\ {\cal{V}}_{\beta}({\cal Z})\ $
 which  corresponds to the root system $\{\beta\}$ of
the Lie superalgebra ${\cal A}_{n,n-1}$ and a super field  $ \Phi \big({\cal Z} \big)\ = \ \varphi (z)\ + i\ \theta \ \psi (z)$,
$$
{\cal{V}}_{_\beta}({\cal Z})=\  \ : e^{i  \beta . \Phi  \big({\cal Z} \big) } :
\eqno(3.1)
$$
In the component formalism, one gets
$$
{\cal{V}}_{_\beta}({\cal Z})={\cal{V}}_{_\beta}^{_b}(z)\ + \ \theta \ {\cal{V}}_{_\beta}^{_f}({z})
\eqno(3.2)
$$
where the bosonic\ ${\cal{V}}_{_\beta}^{_b}(z)$ and fermionic\ ${\cal{V}}_{_\beta}^{_f}(z)$
 components of ${\cal{V}}_{_\beta}({\cal Z})$ are
$$
{\cal{V}}_{_\beta}^{_b}(z)=\  \ : e^{i \ \beta . \varphi  (z) }:
\eqno(3.3)
$$
and
$$
{\cal{V}}_{_\beta}^{_f}(z)=\ \beta . \Psi (z)  : e^{i \ \beta . \varphi  (z) }:
\eqno(3.4)
$$
respectively.We must emphasize here that we will concentrate over  the bozonic vertex operator 
${\cal{V}}_{_\beta}^{_b}(z)$,\ in our advanced calculations.\ By using conformal spin-0  
contraction $\varphi(\underline{z_1)\,\varphi(}z_2)=-\ln \mid z_{12} \mid$,\ 
The standard OPEs are of two forms:
$$
{\cal{V}}^b_{_\beta}(z_1)\,{\cal{V}}^b_{_{\dot{\beta}}}(z_2)\,=(z_{12})^{_{\beta\,\dot{\beta}}}\,
:{\cal{V}}^b_{_\beta}(z_1)\,{\cal{V}}^b_{_{\dot{\beta}}}(z_2):
\eqno(3.5)
$$
and
$$
{\cal{V}}^f_{_\beta}(z_1)\,{\cal{V}}^f_{_{\dot{\beta}}}(z_2)\,={{\beta^2}\over{ z_{12}}}
{\cal{V}}_{_\beta}^b(z_1)\,{\cal{V}}_{_{\dot{\beta}}}^b(z_2)
\eqno(3.6)
$$
One can say that the operators ${\cal{V}}^b_{_\beta}(z)$ and  ${\cal{V}}^f_{_\beta}(z)$
carrie a root $\beta$ .\ From
$$
h_{_j}(z_1)\,{\cal{V}}^b_{\beta}(z_2)=
{{\theta_{_j}}\over {z_{12}}}  {\cal{V}}^b_{_{\beta}}(z_2)\,+\,\cdots
\eqno(3.7)
$$
and
$$
h_{_j}(z_1)\,{\cal{V}}^f_{\beta}(z_2)=
{{\theta_{_j}}\over {z_{12}}}  {\cal{V}}^f_{_{\beta}}(z_2)\,+\,\cdots
\eqno(3.8)
$$
where
$$
\theta_{_j}=
\theta_{_j}(\beta)
\equiv\,
(\beta,\lambda_{_{j}})\,
\eqno(3.9)
$$

The OPEs with the stress-energy tensor $\top(z)$ is
$$
\top(z_1)\,{\cal{V}}^b_{\beta}(z_2)=
{h^b(\beta)\over {z_{12}^2}}  {\cal{V}}^b_{_{\beta}}(z_2) \,+\,
{({\eta_{_b}}^{\beta} {\cal{V}}^b_{_{\beta}})(z_2)\over {z_{12}}} \,+\,\cdots
\eqno(3.10)
$$
where  $h^b(\beta)$ is given by
$$
h^b(\beta)=-\sum_{i=1}^{2 n-1}(-1)^{i+1} \theta_i\theta_{i+1}+ {a_{_0}\over 2}
\sum_{i=1}^{2 n} \theta_i
\eqno(3.11)
$$
this means that the bosonic  vertex operator ${\cal{V}}^b_{_{\beta}}(z)$ is a conformal 
field with  spin $h^b(\beta)$, and
$$
({\eta_{_b}}^{\beta} {\cal{V}}^b_{_{\beta}})(z)
=-\sum_{i=1}^{2 n-1}(-1)^{i+1}\big( \theta_i h_{i+1}(z)+ h_i(z) \theta_{i+1}\big)
\eqno(3.12)
$$
similarly, for the fermionic vertex operator ${\cal{V}}^f_{_{\beta}}(z)$  
$$
\top(z_1)\,{\cal{V}}^f_{\beta}(z_2)=
{{({h_{_f}}^{\beta} {\cal{V}}^f_{_{\beta}})(z_2)}\over{z_{12}^2}} \,+\,
{({\eta_{_f}}^{\beta} {\cal{V}}^b_{_{\beta}})(z_2)\over {z_{12}}} \,+\,\cdots
\eqno(3.13)
$$
where $({h_{_f}}^{\beta} {\cal{V}}^f_{_{\beta}})(z)$
$$
({h_{_f}}^{\beta} {\cal{V}}^f_{_{\beta}})(z)=
{1\over 2}\sum_{i=1}^{2 n-1}(-1)^{i+1}
(\theta_i \psi_{i+1}(z_2)+\psi_{i}(z_2)\theta_{i+1}) {\cal{V}}^b_{_{\beta}}(z_2)
$$
$$
+\sum_{i=1}^{2 n-1}(-1)^{i} \theta_i \theta_{i+1}  {\cal{V}}^f_{_{\beta}}(z_2)
+{{a_{_0}}\over{2}}\sum_{i=1}^{2 n} \theta_i  {\cal{V}}^f_{_{\beta}}(z_2)
\eqno(3.14)
$$
and
$$
({\eta_{_f}}^{\beta} {\cal{V}}^b_{_{\beta}})(z)=
\sum_{i=1}^{2 n-1}(-1)^{i+1}
(\theta_i \partial \psi_{i+1}(z_2)+\partial \psi_{i}(z_2)\theta_{i+1}) {\cal{V}}^b_{_{\beta}})(z_2)
$$
$$
+\sum_{i=1}^{2 n-1}(-1)^{i}
(\theta_i h_{i+1}(z_2)+h_{i}(z_2)\theta_{i+1}) {\cal{V}}^f_{_{\beta}})(z_2)
\eqno(3.15)
$$
From (3.13), we can not define  $h^f(\beta)$ as the conformal spin of
 ${\cal{V}}^f_{_{\beta}}(z)$, under (3.9) definition, since ${\cal{V}}^f_{_{\beta}}(z)$
field does not seem in the r.h.s. of OPE (3.13).

Let us discuss the OPEs between the quasi-primary fields of the super-Casimir ${\cal {WA}}$$_{_n}$ algebra and the bosonic  vertex operator ${\cal{V}}^b_{_{\beta}}(z)$ ,
instead of primary fields since they are very complicated, which are given as in equation (2.12).\ First we write down the OPEs between module-1 components
${\cal J}_{_{1}}({z})$ and  ${\cal G}_{_{{3\over2}}}^{\mp}({z})$ are given by \ :
$$
{\cal{J}}_{1}(z_1)\,{\cal{V}}^b_{\beta}(z_2)=\sum_{i=1}^{2 n}(-1)^{i-1}\theta_i
{{\cal{V}}^b_{_{\beta}}(z_2)\over {z_{12}}} \,+\,\cdots
\eqno(3.16)
$$
and, for fermionic curreents :
$$
{\cal{G}}_{{3\over 2}}^{_+}(z_1)\,{\cal{V}}^b_{\beta}(z_2)=
\Bigg(
-\sum_{i=1}^{2 n-2}(1-t_i) \theta_{i}\psi_{i+1}(z)
+\sum_{i=1}^{2 n} t_i \theta_{i+1}\psi_{i}(z)
 \Bigg)
{{\cal{V}}^b_{_{\beta}}(z_2)\over {z_{12}}} \,+\,\cdots
\eqno(3.17)
$$
$$
{\cal{G}}_{{3\over 2}}^{_+}(z_1)\,{\cal{V}}^b_{\beta}(z_2)=
\Bigg(
\sum_{i=1}^{2 n-2}(1-t_i) \theta_{i+1}\psi_i(z)
-\sum_{i=1}^{2 n} t_i \theta_{i}\psi_{i+1}(z)
 \Bigg)
{{\cal{V}}^b_{_{\beta}}(z_2)\over {z_{12}}} \,+\,\cdots
\eqno(3.18)
$$
\noindent Second the OPEs between module-2 components
${\cal J}_{_{2}}({z})$ ,\  ${\cal G}_{_{{5\over2}}}^{\mp}({z})$, ${\cal T}_{_{3}}({z})$
and the bosonic  vertex operator ${\cal{V}}^b_{_{\beta}}(z)$ are given by \ :

$$
{\cal{J}}_{2}(z_1)\,{\cal{V}}^b_{\beta}(z_2)=
 \Bigg(
a_{_0}^2\sum_{i=1}^{2 n-3} \sum_{j=i+2}^{2 n} (-1)^{i+j} t_i \theta_i \theta_j+
a_{_0}^2\sum_{i=1}^{2 n-2} \sum_{j=i+1}^{2 n} (-1)^{j}(1-t_i)\theta_i \theta_j
$$
$$
-{1\over 2}a_{_0}^3\sum_{i=3}^{2 n} (-1)^{i+1}\theta_i \big((i-1)t_i+(i-2)(1-t_i)\big)
 \Bigg)
{{\cal{V}}^b_{_{\beta}}(z_2)\over {z_{12}^2}}+\cdots
\eqno(3.19)
$$
and, for fermionic curreents :
$$
{\cal{G}}_{{5\over 2}}^{_+}(z_1)\,{\cal{V}}^b_{\beta}(z_2)=
\Bigg(
a_{_0}\sum_{i=1}^{2 n-3} \sum_{j=i+2}^{2 n}(-1)^{j+1} t_i \psi_{i}(z_2)\theta_{i+1}\theta_{j}
+a_{_0}\sum_{i=1}^{2 n-3} \sum_{j=i+1}^{2 n-1}(-1)^{i} (1-t_i)\theta_{i}\theta_{j}\psi_{j+1}(z_2)
$$
$$
+a_{_0}\sum_{i=1}^{2 n-3} \sum_{j=i+2}^{2 n-1}(-1)^{i+1} t_i \theta_{i}\psi_{j}(z_2)\theta_{j+1}
+a_{_0}\sum_{i=1}^{2 n-4} \sum_{j=i+3}^{2 n}(-1)^{j} (1-t_i) \theta_{i}\psi_{j+1}(z_2)\theta_{j }
$$
$$
+{1\over 2}a_{_0}^2\sum_{i=1}^{2 n-2}i(-1)^{i} (1-t_i) \theta_{i}\psi_{i+1}(z_2)
+{1\over 2}a_{_0}^2\sum_{i=1}^{2 n-1}(i-1)(-1)^{i} t_i \psi_{i}(z_2)\theta_{i+1}
 \Bigg)
{{\cal{V}}^b_{_{\beta}}(z_2)\over {z_{12}}^2} + \cdots
\eqno(3.20)
$$
$$
{\cal{G}}_{{5\over 2}}^{_-}(z_1)\,{\cal{V}}^b_{\beta}(z_2)=
\Bigg(
-a_{_0}\sum_{i=1}^{2 n-2} \sum_{j=i+2}^{2 n}(-1)^{j}(1-t_i) \psi_{i}(z_2)\theta_{i+1}\theta_{j}
+a_{_0}\sum_{i=1}^{2 n-3} \sum_{j=i+3}^{2 n-1}(-1)^{j} t_i\theta_{i}\psi_{i+1}(z_2)\theta_{j}
$$
$$
-a_{_0}\sum_{i=1}^{2 n-4}\sum_{j=i+2}^{2 n-2}(-1)^{i}(1-t_i)\theta_{i}\psi_{j}(z_2)\theta_{j+1}
+a_{_0}\sum_{i=1}^{2 n-2} \sum_{j=i+1}^{2 n-1}(-1)^{i} t_i \theta_{i}\theta_{j} \psi_{j+1}(z_2)
$$
$$
-{1\over 2}a_{_0}^2\sum_{i=1}^{2 n-1}(i-1)(-1)^{i} t_i \theta_{i}\psi_{i+1}(z_2)
-{1\over 2}a_{_0}^2\sum_{i=4}^{2 n-2}(i-2)(1- t_i) \psi_{i}(z_2)\theta_{i+1}
 \Bigg)
{{\cal{V}}^b_{_{\beta}}(z_2)\over {z_{12}}^2} + \cdots
\eqno(3.21)
$$
and also finally the last OPE is found to be\ :
$$
{\cal{T}}_{3}(z_1)\,{\cal{V}}^b_{\beta}(z_2)=
- \Bigg(
 a_{_0}\sum_{i=1}^{2 n} \sum_{j=i+3}^{2 n} (-1)^{i+j}\theta_i \theta_{i+1} \theta_j
+a_{_0}\sum_{i=1}^{2 n-3} \sum_{j=i+2}^{2 n-1} (-1)^{i+j}\theta_i \theta_{j} \theta_{j+1}
$$
$$
-a_{_0}^2\sum_{k=3}^{2 n-1} \sum_{i=1}^{2 n-k} t_k\theta_i \theta_{i+k}
+2 a_{_0}^2\sum_{i=1}^{2 n} \sum_{j=i+2}^{2 n} (1-t_i)(1-t_j)\theta_i \theta_{j}
- a_{_0}^2\sum_{i=3}^{2 n-1}(-1)^{i+1} t_i (i-1) \theta_i \theta_{i+1}
$$
$$
- a_{_0}^2\sum_{i=3}^{2 n-2}(-1)^{i+1} (1-t_i) (i-2) \theta_i \theta_{i+1}
+ a_{_0}^3\sum_{i=1}^{2 n}(1- t_i) (i-2) \theta_i
 \Bigg){{\cal{V}}^b_{_{\beta}}(z_2)\over {z_{12}}^3}+ \cdots
\eqno(3.22)
$$
Although the above  advenced calculations was performed in the quasi-primary basis,this
results can be carry on the primary basis.
\vskip 10mm
\noindent{\bf{ Acknowledgments}}
\vskip 5mm
\noindent It is a pleasure to thank K.Ito for helpful informations via e-mail
 communications.


\vskip 5mm
\noindent{\bf References}

\vskip 5mm

\noindent 1. P.\ Bouwknegt and K.\ Schoutens, {\it{Phys. Rep}}. {\bf{223}} (1993) 183 (hep-th/9210010)

\noindent 2. F. Bais, P. Bouwknegt, M. Surridge and K. Schoutens,{\it{ Nucl. Phys}}.
{\bf{B304}} (1988) 348; 371.

\noindent 3. V.A. Fateev and S.L. Luk'yanov, {\it{Sov. Sci. Rev. A Phys}}.15/2 (1990)

\noindent 4. A. B. Zamolodchikov, {\it{Theor. Math. Phys}}. {\bf{65}}, 1205 (1986).

\noindent 5. R.\ Blumenhagen,\ W.\ Eholzer, A.\ Honecker R.\ Hubel  and
K.\ Hornfeck,\ {\it{Int.\ J.\ Mod.\ Phys}}.{\bf{A10}}, 2367 (1995).

\noindent 6. F. Bais, P. Bouwknegt, M. Surridge and K. Schoutens,{\it{ Nucl. Phys}}.
{\bf{B304}} (1988) 348; 371.

\noindent 7. J.M. Figueroa-O'Farrill, S. Schrans and K. Thielemans, "On the
Casimir algebra of $B_2$", {\it{Phys. Lett}}. {\bf{263B}} (1991) 378.

\noindent 8. C.\ "Ahn c=5/2 Free Fermion Model of ${WB}_{_{2}}$ Algebra",
{\it{Int. J. Mod. Phys}}. {\bf{A 6}} (1991) 3467.

\noindent 9. G.M.T. Watts, "$\cal{WB}$ Algebra Representation Theory",
{\it{Nucl. Phys}}. {\bf{B339}} (1990) 177; "$\cal{W}$- Algebras and Coset Models",
{\it{Phys. Lett}}. {\bf{245B}} (1990)65, DAMTP-90-37.

\noindent 10. H.T. Ozer, {\it{Int.\ J.\ Mod.\ Phys}}.{\bf{A14}}, 469 (1999).

\noindent 11.  L.J.Romans, " The $\cal{N}$=2 Super ${W}_{_{3}}$- Algebra "
{\it{Nucl. Phys.}}. {\bf{369B}} (1992) 403.

\noindent 12. D.Nemeschansky and S.Yankielowicz, " The $\cal{N}$=2 ${W}$ -Algebras,
Kazama-Suzuki models and Drinfel`d- Sokolov reduction," USC Preprint USC-91/005(1991).

\noindent 13.  K.Ito, " $\cal{N}$=2 Superconformal ${CP}_{_{n}}$ Model "
{\it{Nucl. Phys.}}. {\bf{370B}} (1992) 123.

\noindent 14.  K.Ito, " Free Field Realization  of $\cal{N}$=2
Super ${W}_{_{3}}$ -Algebra " {\it{Phys. Lett}}. {\bf{304B}} (1993) 271.

\noindent 15. K.Ito, {\it{Nucl. Phys.}} {\bf{370B}} (1992) 123.

\noindent 16. K.Ito, {\it{Phys. Lett.}} {\bf{304B}} (1993) 271.

\noindent 17. K.\ Thielemans,"A  ${Mathematica^{TM}}$ package for computing
operator product expansions (OPEdefs 3.1)",\ Theoretical Phys.Group,\ Imperial
College,\ London(UK).

\noindent 18.\ S. Wolfram, ${Mathematica^{TM}}$, (Addison-Wesley,1990).
\vskip 5mm
\end

\noindent{\bf Appendix}
$$
{\cal W}_{_2}({\cal Z})=
-\,\sum _{i=1}^{n+1}\, \sum _{j=i+2}^{2 n-1}\,({\cal H}_{_i} {\cal H}_{_{i+1}} {\cal H}_j {\cal H}_{_{j+1}} )({\cal Z})
+a_0 \sum_{i=1}^{2n-2}\, \sum _{j=i+2}^{2 n}\, (-1)^j ({\cal H}_{_i} {\cal H}_{_{i+1}} {\cal D} {\cal H}_{_j})({\cal Z})
$$
$$
+a_0 \sum_{i=1}^{2n-2}\, \sum_{j=i+1}^{2 n-1}\, (-1)^i ({\cal D} {\cal H}_{_i} {\cal H}_{_j} {\cal H}_{_{j+1}})({\cal Z})
+{1\over2} a0^2 \sum_{i=1}^{2n-1}\,  (i+t_i-2) ({\cal H}_{_i} {\cal D}^2 {\cal H}_{_{i+1}})({\cal Z})
$$
$$
+    a_0^2 \sum_{i=1}^{2n-3}\, \sum_{j=i+2}^{2 n}\,(-1)^{i+j} t_i ({\cal D} {\cal H}_{_i} {\cal D} {\cal H}_{_j}) ({\cal Z})
+    a_0^2 \sum_{i=1}^{2n-2}\, \sum_{j=i+1}^{2 n}\,(-1)^{i+j} (1-t_i) ({\cal D} {\cal H}_{_i} {\cal D} {\cal H}_{_j})({\cal Z})
$$
$$
+{1\over2} a_0^2 \sum_{i=1}^{2n-1}\, (t_i-i) ({\cal D}^2 {\cal H}_{_i} {\cal H}_{_{i+1}})({\cal Z})
-{1\over2} a_0^3 \sum_{i=1}^{2n}\,(-1)^i (-2+i+t_i) ({\cal D}^3 {\cal H}_{_i})({\cal Z})
$$
$$
{\cal W}_{_{5\over2}}({\cal Z})=
\sum_{i=1}^{2n-4}\, \sum_{j=i+2}^{2 n-1}\,(t_{_j}-1) ({\cal H}_{_i} {\cal H}_{_{i+1}} {\cal H}_{_j} {\cal D} {\cal H}_{_{j+1}})({\cal Z})
+\sum_{i=1}^{2n-3}\, \sum_{j=i+2}^{2 n-1}\,t_{_j} ({\cal H}_{_i} {\cal H}_{_{i+1}} {\cal D} {\cal H}_{_j} {\cal H}_{_{j+1}})({\cal Z})
$$
$$
+\sum_{i=1}^{2n-3}\, \sum_{j=i+2}^{2 n-1}\,(t_{_i}-1) ({\cal H}_{_i} {\cal D} {\cal H}_{_{i+1}} {\cal H}_{_j} {\cal H}_{_{j+1}})({\cal Z})
+\sum_{i=1}^{2n-3}\, \sum_{j=i+2}^{2 n-1}\,t_{_i} ({\cal D} {\cal H}_{_i} {\cal H}_{_{i+1}} {\cal H}_{_j} {\cal H}_{_{j+1}})({\cal Z})
$$
$$
+a_0 \big{(}
\sum_{i=1}^{2n-3}\, \sum_{j=i+1}^{2 n-1}\, t_{_i}({\cal H}_{_{2 n-j}} {\cal H}_{_{2 n-j+1}} {\cal D}^2 {\cal H}_{_{2 n-i+1}})({\cal Z})
+\sum_{i=1}^{2n-3}\, \sum_{j=i+2}^{2 n-1}\, t_{_j}({\cal D}^2  {\cal H}_{_{2 n -j+1}} {\cal H}_{_{2 n-i}} {\cal H}_{_{2 n +1-i}})({\cal Z})
$$
$$
-\sum_{i=1}^{2n-3}\, \sum_{j=i+1}^{2 n-2}\, (-1)^i (t_{_j}-1) ({\cal H}_{_{2 n-j}} {\cal D} {\cal H}_{_{2 n +1-j}} {\cal D} {\cal H}_{_{2 n +1-i}})({\cal Z})
-\sum_{i=1}^{2n-3}\, \sum_{j=i+2}^{2 n}\, (-1)^j t_{_i} ({\cal D} {\cal H}_{_{2 n +1-j}} {\cal D} {\cal H}_{_{2 n-i}} {\cal H}_{_{2 n +1-i}})({\cal Z})
$$
$$
-\sum_{i=1}^{2n-3}\, \sum_{j=i+2}^{2 n-1}\, (-1)^i t_{_j} ({\cal D} {\cal H}_{_{2 n -j}} {\cal H}_{_{2 n +1-j}} {\cal D} {\cal H}_{_{2 n +1-i}})({\cal Z})
-\sum_{i=1}^{2n-4}\, \sum_{j=i+3}^{2 n}\, (-1)^j (t_{_i}-1)( {\cal D} {\cal H}_{_{2 n +1-j}} {\cal H}_{_{2 n -i}} {\cal D} {\cal H}_{_{2 n+1-i}})({\cal Z})
\big{)}
$$
$$
+a_0^2 \sum_{i=1}^{2n-3}\, \sum_{j=i+2}^{2 n}\,(-1)^{i+j} t_{_i} ({\cal D} {\cal H}_{_{2 n+1-j}} {\cal D}^2 {\cal H}_{_{2 n+1-i}})({\cal Z})
+{1\over2} a_0^2 \sum_{i=1}^{2n-3}\,(1+i-2 n) t_{_i} ({\cal D} {\cal H}_{_{2 n-i}}{\cal D}^2  {\cal H}_{_{2 n+1-i}})({\cal Z})
$$
$$
+ {1\over2} a_0^2 \sum_{i=1}^{2n-4}\,  (-2-i+2 n) (1-t_{_i}) ({\cal D}^2 {\cal H}_{_{2 n-i}}{\cal D} {\cal H}_{_{2 n+1-i}})({\cal Z})
+           a_0^2 \sum_{i=1}^{2n-3}\, \sum_{j=i+2}^{2 n-1}\, (-1)^{i+j} t_{_i} ({\cal D}^2 {\cal H}_{_{2 n+1-j}} {\cal D} {\cal H}_{_{2 n +1-i}})({\cal Z})
$$
$$
- {1\over2} a_0^2 \sum_{i=1}^{2n-1}\,  (-1-i+2 n) t_{_i} ({\cal D}^3 {\cal H}_{_{2 n-i}} {\cal H}_{_{2 n+1-i}})({\cal Z})
+ {1\over2} a_0^2 \sum_{i=1}^{2n-4}\,  (2+i-2 n) (t_{_i}-1) ({\cal H}_{_{2 n-i}} {\cal D}^3 {\cal H}_{_{2 n+1-i}})({\cal Z})
$$
$$
- {1\over2} a_0^3 \sum_{i=1}^{2n-3}\,  (-1-i+2 n) t_{_i} {\cal D}^4 {\cal H}_{_{2 n+1-i}}({\cal Z})
$$
\par\vfill\eject

$$
{\cal W}_{_2}({\cal Z})=
      \sum_{i=1}^{2n-3}\, \sum_{j=i+2}^{2 n-1}\,(\psi_{_{i}} \psi_{_{i+1}} \psi_{_{j}} \psi_{_{j+1}} )(z)
-a_{_{0}}   \sum_{i=1}^{2n-2}\, \sum_{j=i+2}^{2 n}\,(-1)^j (\psi_{_{i}} \psi_{_{i+1}} h_{_{j}})(z)
-a_{_{0}}   \sum_{i=1}^{2n-2}\, \sum_{j=i+1}^{2 n-1}\,(-1)^i (h_{_{i}} \psi_{_{j}} \psi_{_{j+1}})(z)
$$
$$
+a_{_{0}}^2 \sum_{i=1}^{2n-3}\, \sum_{j=i+2}^{2 n}\,(-1)^{i+j} t_{_{i}} (h_{_{i}} h_{_{j}})(z)
+a_{_{0}}^2 \sum_{i=1}^{2n-2}\, \sum_{j=i+1}^{2 n}\,(-1)^j (1-t_{_{i}}) (h_{_{i}} h_{_{j}})(z)
+a_{_{0}}^2 \sum_{i=1}^{n-3}\, i (\psi2_{_{ i+1}} \psi'_{_{2 i+2}}
$$
$$
+a_{_{0}}^2 \sum_{i=1}^{n-2}\, i (\psi2_{_{ i+2}} \psi'_{_{2 i+3}})(z)
+a_{_{0}}^2 \sum_{i=1}^{n-1}\, i (\psi'_{_{2 i}} \psi_{_{2 i+1}})(z)
+a_{_{0}}^2 \sum_{i=1}^{n-1}\, i (\psi'_{_{2 i+1}} \psi_{_{2 i+2}})(z)
+a_{_{0}}^3 \sum_{i=1}^{n-1}\, i h'_{_{2 i+1}}(z)
-a_{_{0}}^3 \sum_{i=1}^{n-1}\, i h'_{_{2 i+2}}(z)
$$
$$
{\cal W}^{-}(z)=
+\sum_{i=1}^{2n-4}\,\sum_{j=i+2}^{2 n-1}\,(1-t_{_{j}}) (\psi_{_{i}} \psi_{_{i+1}} \psi_{_{j }} h_{_{j+1}})(z)
+\sum_{i=1}^{2n-3}\,\sum_{j=i+2}^{2 n-1}\,(-1)^j t_{_{j}} (\psi_{_{i}} \psi_{_{i+1}} h_{_{j}} \psi_{_{j+1}})(z)
$$
$$
+\sum_{i=1}^{2n-3}\, \sum_{j=i+2}^{2 n-1}\,(1-t_{_{i}}) (\psi_{_{i}} h_{_{i+1}} \psi_{_{j}} \psi_{_{j+1}})(z)
+\sum_{i=1}^{2n-3}\, \sum_{j=i+2}^{2 n-1}\,(-1)^i t_{_{i}} (h_{_{i}} \psi_{_{i+1}} \psi_{_{j}} \psi_{_{j+1}})(z)
$$
$$
-a_{_{0}} \sum_{i=1}^{2n-2}\, \sum_{j=i+2}^{2 n}\,(1-t_{_{j}}) (\psi_{_{i}} \psi_{_{i+1}} \psi'_{_{j}})(z)
-a_{_{0}} \sum_{i=1}^{2n-2}\, \sum_{j=i+2}^{2 n}\,(-1)^j (1-t_{_{i}}) (\psi_{_{i}} h_{_{i+1}} h_{_{j}})(z)
$$
$$
+a_{_{0}} \sum_{i=1}^{2n-3}\, \sum_{j=i+2}^{2 n-1}\,(-1)^{i+j} t_{_{j}} (h_{_{2 n -j}} \psi_{_{2 n +1-j}} h_{_{2 n +1-i}})(z)
+a_{_{0}} \sum_{i=1}^{2n-4}\, \sum_{j=i+3}^{2 n}\,(-1)^{i+j} (1-t_{_{i}}) (h_{_{2 n +1-j}} \psi_{_{2 n -i}} h_{_{2 n+1-i}})(z)
$$
$$
-a_{_{0}} \sum_{i=1}^{2n-3}\, \sum_{j=i+2}^{2 n}\,(-1)^j t_{_{i}}(  h_{_{2 n +1-j}} h_{_{2 n-i}} \psi_{_{2 n +1-i}})(z)
-a_{_{0}} \sum_{i=1}^{2n-3}\, \sum_{j=i+2}^{2 n-1}\, t_{_{j}} (\psi'_{_{2 n -j+1}}  \psi_{_{2 n-i}} \psi_{_{2 n +1-i}})(z)
$$
$$
+a_{_{0}}^2 \sum_{i=1}^{2n-3}\, \sum_{j=i+2}^{2 n}\,(-1)^{i+j} t_{_{i}} (h_{_{2 n+1-j}} \psi'_{_{2 n+1-i}})(z)
+{1\over2} a_{_{0}}^2  \sum_{i=1}^{2n-3}\,(1+i-2 n)  t_{_{i}} (h_{_{2 n-i}} \psi'_{_{2 n+1-i}})(z)
$$
$$
+a_{_{0}}^2 \sum_{i=1}^{2n-3}\, \sum_{j=i+2}^{2 n-1}\,(-1)^{i+j} t_{_{j}} (\psi'_{_{2 n+1-j}} h_{_{2 n +1-i}})(z)
+{1\over2} a_{_{0}}^2  \sum_{i=1}^{2n-4}\, \sum_{j=i+1}^{2 n-1}\,(-2-i+2 n) (1-t_{_{i}}) (\psi'_{_{2 n-i}} h_{_{2 n+1-i}})(z)
$$
$$
-{1\over2} a_{_{0}}^2  \sum_{i=1}^{2n-1}\, \sum_{j=i+1}^{2 n-1}\,(-1-i+2 n) t_{_{i}} (h'_{_{2 n-i}} \psi_{_{2 n+1-i}})(z)
-{1\over2} a_{_{0}}^2  \sum_{i=1}^{2n-4}\, \sum_{j=i+1}^{2 n-1}\,(2+i-2 n)  (1-t_{_{i}}) (\psi_{_{2 n-i}} h'_{_{2 n+1-i}})(z)
$$
$$
-{1\over2} a_{_{0}}^3  \sum_{i=1}^{2n-3}\,(-1-i+2 n) t_{_{i}} \psi''_{_{2 n+1-i}}(z)
$$
$$
{\cal W}^{+}(z)=
{\cal W}^{-}(z)
\Bigg{[}
\matrix {h_{_{i}}\rightarrow h_{_{2 n+1-i}}\cr
\psi_{_{i}}\rightarrow +\psi_{_{2 n+1-i}}}
 \Bigg{]}
-R(z)
$$
$$
R(z)=
-a_{_{0}}^2 \sum_{i=1}^{n}\,(n-1) (\psi'_{_{2 i-1}} h_{_{2 i}})(z)
-a_{_{0}}^2 \sum_{i=1}^{n}\,(n-1) (\psi_{_{2 i-1}} h'_{_{2 i}})(z)
+a_{_{0}}^2 \sum_{i=1}^{n-1}\,(n-1) (h_{_{2 i}} \psi'_{_{2 i+1}})(z)
$$
$$
+a_{_{0}}^2 \sum_{i=1}^{n-1}\,(n-1) (h'_{_{2 i}} \psi_{_{2 i+1}})(z)
-a_{_{0}}^3 \sum_{i=1}^{n}\,(n-1) \psi''_{_{2 i-1}}(z)
$$
$$
{\cal W}_{_3}({\cal Z})=
 \sum_{i=1}^{2n-3}\, (-1)^i                                                 ( h_{_{i}} \psi_{_{i+1}} \psi_{_{i+2}} h_{_{i+3}})(z)
+ \sum_{i=1}^{2n-3}\, \sum_{j=i+4}^{2 n-1}\ (-1)^i t_{_{i}} t_{_{j}}         ( h_{_{i}} \psi_{_{i+1}} \psi_{_{j}} h_{_{j+1}})(z)
$$
$$
+ \sum_{i=1}^{2n-3}\, \sum_{j=i+3}^{2 n-1}\ (-1)^i (1-t_{_{i}}) (1-t_{_{j}}) ( h_{_{i}} \psi_{_{i+1}} \psi_{_{j}} h_{_{j+1}})(z)
+ \sum_{i=1}^{2n-4}\, \sum_{j=i+3}^{2 n-2}\ (-1)^j t_{_{i}} (1-t_{_{j}})     ( h_{_{i}} \psi_{_{i+1}} h_{_{j}} \psi_{_{j+1}})(z)
$$
$$
- \sum_{i=1}^{2n-4}\, \sum_{j=i+3}^{2 n-1}\ (1-t_{_{i}}) t_{_{j}}            ( h_{_{i}} \psi_{_{i+1}} h_{_{j}} \psi_{_{j+1}})(z)
+ \sum_{i=1}^{2n-3}\, \sum_{j=i+2}^{2 n-1}\ (-1)^j                           ( \psi_{_{i}} \psi_{_{i+1}} h_{_{j}} h_{_{j+1}})(z)
$$
$$
- \sum_{i=1}^{2n-3}\, \sum_{j=i+1}^{2 n-1}\ (-1)^i t_{_{i}} t_{_{j}}         ( \psi_{_{i}} h_{_{i+1}} h_{_{j}} \psi_{_{j+1}})(z)
- \sum_{i=1}^{2n-3}\, \sum_{j=i+1}^{2 n-1}\ (-1)^j (1-t_{_{i}}) (1-t_{_{j}}) ( \psi_{_{i}} h_{_{i+1}} h_{_{j}} \psi_{_{j+1}})(z)
$$
$$
+ \sum_{i=1}^{2n-4}\, \sum_{j=i+3}^{2 n-2}\ (-1)^i t_{_{i}} (1-t_{_{j}})     ( \psi_{_{i}} h_{_{i+1}} \psi_{_{j}} h_{_{j+1}})(z)
+ \sum_{i=1}^{2n-4}\, \sum_{j=i+3}^{2 n-1}\ (-1)^i (1-t_{_{i}}) t_{_{j}}     ( \psi_{_{i}} h_{_{i+1}} \psi_{_{j}} h_{_{j+1}})(z)
$$
$$
+ \sum_{i=1}^{2n-3}\, \sum_{j=i+2}^{2 n-1}\ (-1)^i                           ( h_{_{i}} h_{_{i+1}} \psi_{_{j}} \psi_{_{j+1}})(z)
+ \sum_{i=1}^{2n-1}\, \sum_{j=i+2}^{2 n-1}\  t_{_{i}}                        ( \psi'_{_{i}}  \psi_{_{i+1}} \psi_{_{j}} \psi_{_{j+1}})(z)
$$
$$
+ \sum_{i=1}^{2n-4}\, \sum_{j=i+2}^{2 n-1}\  (1-t_{_{i}})                    ( \psi_{_{i}} \psi_{_{i+1'}} \psi_{_{j}} \psi_{_{j+1}})(z)
+ \sum_{i=1}^{2n-3}\, \sum_{j=i+2}^{2 n-1}\  t_{_{j}}                        ( \psi_{_{i}} \psi_{_{i+1}} \psi'_{_{j}} \psi_{_{j+1}})(z)
 $$
$$
+ \sum_{i=1}^{2n-4}\, \sum_{j=i+2}^{2 n-1}\  (1-t_{_{j}})                    ( \psi_{_{i}} \psi_{_{i+1}} \psi_{_{j}} \psi'_{_{j+1}})(z)
$$
$$
-a_{_0} \Big{(}
\sum_{i=1}^{2n}\,\sum_{j=i+3}^{2 n}\,t_{_{i}} (1-t_{_{j}}) (\psi_{_{i}} \psi_{_{i+1}} h'_{_{j}})(z)
-\sum_{i=1}^{2n}\,\sum_{j=i+2}^{2 n}\,(1-t_{_{i}}) (1-t_{_{j}})  (\psi_{_{i}}  \psi_{_{i+1}}   h'_{_{j}}  )(z)
+\sum_{k=3}^{2n-1}\, \sum_{i=1  }^{2 n-k}\,  t_{_{k}}                  (\psi_{_{i}}  h_{_{i+1}}   \psi'_{_{i+k}})(z)
$$
$$
- \sum_{i=1}^{2n  }\, \sum_{j=i+2}^{2 n  }\, (-1)^{_j}  (1-t_{_{i}})       (\psi_{_{i}}  \psi'_{_{i+1}}  h_{_{j}}   )(z)
- \sum_{i=1}^{2n-2}\, \sum_{j=i+3}^{2 n  }\, (-1)^{_{i+j}}                 (h_{_{i}}  h_{_{i+1}}   h_{_{j}}   )(z)
- \sum_{i=1}^{2n-3}\, \sum_{j=i+2}^{2 n-1}\, (-1)^{_{i+j}}                 (h_{_{i}}  h_{_{j}}     h_{_{j+1}} )(z)
$$
$$
- \sum_{i=1}^{2n-2}\, \sum_{j=i+1}^{2 n-1}\, (-1)^{_i} t_{_{i}} t_{_{j}} (h_{_{i}}  \psi'_{_{j}}    \psi_{_{j+1}} )(z)
- \sum_{i=1}^{2n-2}\, \sum_{j=i+1}^{2 n-1}\, (1-t_{_{i}}) t_{_{j}}      (h_{_{i}}  \psi'_{_{j}}    \psi_{_{j+1}} )(z)
- \sum_{i=1}^{2n  }\, \sum_{j=i+4}^{2 n  }\, t_{_{i}} t_{_{j}}          (h_{_{i}}  \psi_{_{i+1}}   \psi'_{_{j}}  )(z)
$$
$$
+ \sum_{i=1}^{2n  } \sum_{j=i+2}^{2 n-2} t_{_{i}} (1-t_{_{j}})      (h_{_{i}}  \psi_{_{j}}     \psi'_{_{j+1}})(z)
- \sum_{i=1}^{2n  } \sum_{j=i+2}^{2 n-2} (1-t_{_{i}}) (1-t_{_{j}})  (h_{_{i}}  \psi_{_{j}}     \psi'_{_{j+1}})(z)
- \sum_{i=1}^{2n-2}\,                        (1-t_{_{i}})               (\psi'_{_{i}} \psi_{_{i+1}}   h_{_{i+2}} )(z)
$$
$$
- \sum_{i=1}^{2n-2}\, \sum_{j=i+2}^{2 n  }\, t_{_{i}} (1-t_{_{j}})      (\psi'_{_{i}} \psi_{_{i+1}}   h_{_{j}}   )(z)
+ \sum_{k=4}^{2n-2}\sum_{i=1  }^{2 n-k-1} t_{_{i}} (1-t_{_{k}})      (\psi'_{_{i}} \psi_{_{i+1}}   h_{_{i+k}} )(z)
$$
$$
- \sum_{i=1}^{2n  }\, \sum_{j=i+2}^{2 n  }\, (1-t_{_{i}}) (1-t_{_{j}})  (h_{_{i}}  \psi_{_{i+1}}   \psi'_{_{j}}  )(z)
- \sum_{k=2}^{2n-2}\sum_{i=1  }^{2 n-k-1} t_{_{k}}                   (\psi'_{_{i}} \psi_{_{i+k}}   h_{_{i+k+1}})(z)
$$
$$
+ \sum_{k=1}^{2n-1}\sum_{i=1  }^{2 n-k-1} (1-t_{_{k}})               (\psi'_{_{i}} h_{_{i+k}}   \psi_{_{i+k+1}})(z)
- \sum_{k=1}^{2n-2}\sum_{i=1  }^{2 n-k-1} (1-t_{_{i}})               (h'_{_{i}} \psi_{_{i+k}}   \psi_{_{i+k+1}})(z)
\Big{)}
$$
$$
+a_{_0}^2 \Big{(}            \sum_{i=1}^{2n-2}\, \sum_{j=i+1}^{2 n}\, (1-t_{_{i}}) (1-t_{_{j}}) (h_{_{i}}    h'_{_{j}} )(z)
-             \sum_{i=1}^{2n-1}\, \sum_{j=i+3}^{2 n}\, t_{_{i}} (1-t_{_{j}})     (h_{_{i}}    h'_{_{j}} )(z)
-  {{1\over2}}\sum_{i=1}^{2n-1}\, t_{_{i}} (i-1)(h_{_{i}} h'_{_{i+1}} )(z)
$$
$$
+  {{1\over2}}\sum_{i=1}^{2n-2}\, (1-t_{_{i}})   (i-2)(h_{_{i}}    h'_{_{i+1}} )(z)
+  {{1\over2}}\sum_{i=1}^{2n-2}\, (1-t_{_{i}})   (i-2)(h'_{_{i}}   h_{_{i+1}}  )(z)
-  {{1\over2}}\sum_{i=1}^{2n-1}\, t_{_{i}}      (i-1)    (h'_{_{i}}   h_{_{i+1}}  )(z)
$$
$$
+             \sum_{i=1}^{2n-2}\, \sum_{j=i+2}^{2 n}\,   (1-t_{_{i}}) (1-t_{_{j}}) (h'_{_{i}}   h_{_{j}}  )(z)
-             \sum_{i=1}^{2n-4}\, \sum_{j=i+3}^{2 n-1}\,       (1-t_{_{i}}) t_{_{j}}     (h'_{_{i}}   h_{_{j}}  )(z)
+             \sum_{i=1}^{2n-1}\, \sum_{j=i+2}^{2 n}\,       t_{_{i}} (1-t_{_{j}})     (\psi'_{_{i}}   \psi'_{_{j}} )(z)
$$
$$
+  {{1\over2}}\sum_{i=1}^{2n-1}\,                         t_{_{i}}   (i-1)        (\psi'_{_{i}}   \psi'_{_{i+1}} )(z)
-             \sum_{i=1}^{2n-1}\, \sum_{j=i+3}^{2 n}\,                        (1-t_{_{i}}) t_{_{j}}     (\psi'_{_{i}}   \psi_{_{j}}  )(z)
+  {{1\over2}}\sum_{i=1}^{2n-2}\,                        (1-t_{_{i}})    (i-2)   (\psi'_{_{i}}   \psi'_{_{i+1}} )(z)
$$
$$
+  {{1\over2}}\sum_{i=1}^{2n-1}\,                         t_{_{i}}    (i-1)       (\psi''_{_{i}}  \psi_{_{i+1}}  )(z)
+  {{1\over2}}\sum_{i=1}^{2n-1}\,                        (1-t_{_{i}})    (i-2)   (\psi_{_{i}}    \psi''_{_{i+1}})(z)
\Big{)}
+a_{_0}^3 \,\, {{1\over2}}\sum_{i=1}^{2n}\,(1-t_{_{i}})   (2 - i) h''_{_{i}} (z)
$$
\end